\documentclass[aps,pre,twocolumn,groupedaddress]{revtex4-1}

\usepackage{amsmath, color}
\usepackage{graphicx}

\begin{document}


\title{The formation and arrangement of pits by a corrosive gas }


\author{James Burridge}
\affiliation{Department of Mathematics, University of Portsmouth, Portsmouth PO1 3HF, United Kingdom}

\author{Robert Inkpen}
\affiliation{Department of Geography, University of Portsmouth, Portsmouth PO1 3HE, United Kingdom}


\date{\today}

\begin{abstract}
When corroding or otherwise aggressive particles are incident on a surface, pits can form. For example, under certain circumstances rock surfaces that are exposed to salts can form regular tessellating patterns of pits known as ``tafoni''. We introduce a simple lattice model in which a gas of corrosive particles, described by a discrete convection diffusion equation, drifts onto a surface. Each gas particle has a fixed probability of being absorbed and causing damage at each contact. The surface is represented by a lattice of strength numbers which reduce after each absorbtion event, with sites being removed when their strength becomes negative. The model generates regular formations of pits, with each pit having a characteristic trapezoidal geometry determined by the particle bias,  absorbtion probability and surface strength. The formation of this geometry may be understood in terms of a first order partial differential equation. By viewing pits as particle funnels, we are able to relate the gradient of pit walls to absorbtion probability and particle bias.
\end{abstract}


\maketitle

\section{Introduction}
\label{intro}

Pits forming in clusters on the surfaces of rocks have been studied for over a century \cite{Must82}. Pitting corrosion is also one of the major damage mechanisms in metals and other materials used in engineering structures \cite{Bur04}. In both rock and metal, the processes which lead to pit formation involve multiple physical phenomena. They have in common that corrosive, or otherwise aggressive material must be transported onto the surface. With this in mind, we investigate pit formation using the simplest possible model that includes a transport process, and for simplicity we refer to the damage caused by particles, which might in practice not be chemical, as corrosion. Our aim is to discover what structures are formed on a surface when corrosive or otherwise aggressive particles are biased toward it.

\begin{figure}
\includegraphics[width=6 cm]{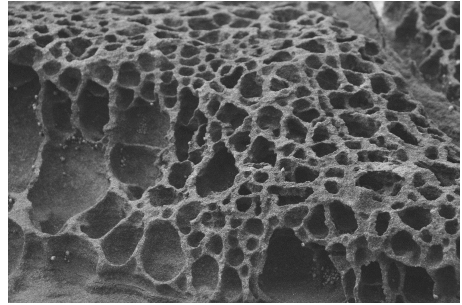}
\caption{ Image of tafoni from Louttit Bay, near Lorne, Victoria, Australia. \label{tafoni}}
\end{figure}

Pit formation is of importance in both geology and engineering. For example, geologists are interested in rock forms created by regular clustering of pits. These occur in many climatic zones including coasts and deserts, and in different lithologies \cite{Must82,Turk04}. An example is shown in Figure \ref{tafoni}. Such formations have been labelled as ``cavernous'', ``alveolar'', ``honeycomb'' and most commonly ``tafoni''. We adopt this latter term, the plural of ``tafone'', a single pit. There is a general consensus that they are created by salt weathering \cite{Must82,Cook93,Mats91}, although some chemical weathering processes \cite{Conc87} and biological agents \cite{Must10} have been suggested as being significant. Salt crystallization and expansion produce stresses in the rock that result in erosion but why regular pits are spontaneously formed rather than merely surface lowering is not yet clear \cite{Turk04,Vil05}. Recent work has also highlighted the importance of a rock's characteristics, for example its strength, porosity or surface harness, as controlling factors \cite{McB04,Mol12}.

Theories for the mechanism driving tafoni growth include hardening of the top layer of rock \cite{Mott94}, softening of the rock core by chemical processes \cite{Conc85}, and wind acceleration in the cavity \cite{Rod99}. However, it has been noted that there is no generally agreed classification of the forms that tafoni can take, or of the formation processes \cite{Turk04,Vil05}, many of which have yet to be modelled mathematically. One proposed mechanism that has been modelled mathematically is the evolution of a single pit driven by the migration and crystallization of salts due to cycles of wetting and drying \cite{Hui04}. Single tafone are theoretically shown to grow as a result of excess salt crystallisation in regions with low evaporation rates, deeper in the pits. In addition to this physical model, a phenomenological approach has been introduced \cite{Sun11}, where a simple functional form is hypothesized for the relationship between the rate at which pits deepen and their age. The form is motivated by viewing the development of tafoni as a process involving positive and negative feedback between pit shape and growth rate. Deepening occurs rapidly at first before reaching a critical depth after which it slows. It appears that no mathematical model exists in which the regular formations of pits seen in nature (see Figure \ref{tafoni}) arise spontaneously.

In engineering applications, pit corrosion in metals is one of the most difficult corrosion mechanisms to manage. Perhaps due to its economic importance, metal pitting has been the subject of a great deal more mathematical research than the formation of tafoni. Pits in metal surfaces are believed to propagate due to the concentration of chlorides inside them \cite{Bur04}. Pits originate as tiny irregular nucleation sites, which can then either stop growing or propagate rapidly \cite{Bur04}. Lattice gas cellular automata are commonly used to model pit formation \cite{Mal05, Pid08, Staf13}, and these models have captured the initiation and propagation of pits, however analytical results on pit geometry or growth rate have yet to be found.

In our model we will see that pits form when fluctuations in the corrosion process create depressions of sufficient depth to collect proportionally more corrosive particles than neighboring regions. These collected particles then cause the depression to deepen relative to the rest of the surface forming a pit. Despite its simplicity, the model shares a common feature with the wetting and drying process \cite{Hui04} and the phenomenological model \cite{Sun11} in that pit growth is a consequence of positive feedback between shape and growth rate: deeper pits initially grow faster. We will also see that pits reach a stable depth, mimicking the negative feedback reported phenomenologically. In addition, as we will see, the model spontaneously produces clusters of pits creating formations of striking regularity. The model is not a complete description of any particular pitting phenomenon. However its simplicity and tractability permit us to gain insight into how regular pits can form, in general.

\section{The model}
\label{modelSect}

\begin{figure}
\includegraphics[width=5.5 cm]{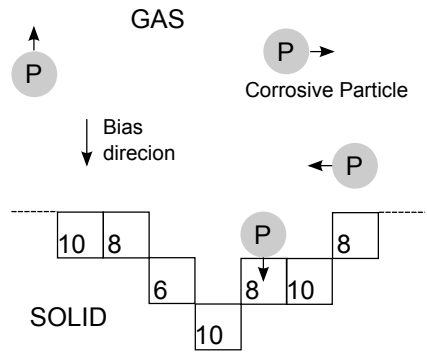}
\caption{Schematic representation of the corrosion process on a two dimensional lattice. The numbers in the surface of the solid matrix represent the strength of the solid in that location. The numbers are reduced when a corrosive particle is absorbed by the surface. The motion of corrosive particles is biased toward the surface. \label{model}}
\end{figure}

Our model is defined in discrete time on an integer lattice, the sites of which can be either gas or solid. The interface between the two regions defines the surface of the solid. We consider lattices of two and three dimensions corresponding to surfaces of dimensions one or two, but our main focus is on one dimensional surfaces. In common with studies of pitting in metals \cite{Pid08,Staf13} we begin by considering a simple model of surface corrosion by discrete particles. In the gas region these perform random walks biased in the direction of the surface. This mimics, in the case of salt weathering of rocks, salt particles being carried onto a rock surface by wind or sea spray. Initially, all solid sites are given a strength number. If a corrosive particle occupies a site adjacent to the surface and its next randomly chosen step would take it into the surface, it is absorbed with probability $p_c$ - the ``corrosion probability''.  Otherwise it remains in its current position. Absorbtion of a particle reduces the strength of the surface site by a fixed quantity. The process is illustrated in Figure \ref{model}.

Although our model is based on the process just described, rather than modelling individual particles, we model the density of a population of particles. In doing so we neglect fluctuations in the corrosion rate over the surface which arise from the discrete nature of the particles. These fluctuations would have the effect of roughening the surface on short length scales, with regular pit formations dominating at larger scales, preventing the formation of a fractal surface  \cite{Bar95}. We arrive at an evolution equation for the particle density by noting that the probability mass function for a single random walking particle evolves according to a discrete time and space master equation \cite{Kamp} with surface boundary condition corresponding to an absorbtion probability $p_c$. If we neglect collisions and exclusion constraints between particles then the master equation also describes the evolution of the density of a population of particles.  If $\phi(\vec{x},t)$ is the particle density at gas site $\vec{x}$ at time $t$, then this evolution equation is:
\begin{align}
\nonumber
\phi(\vec{x},t+1) & = \phi(\vec{x},t) + \sum_{\vec{y} \in G(\vec{x})}  p(\vec{y} \rightarrow \vec{x}) \phi(\vec{y},t) \\
\nonumber
& - \sum_{\vec{y} \in G(\vec{x})}  p(\vec{x} \rightarrow \vec{y}) \phi(\vec{x},t) \\
& - p_c \sum_{\vec{y} \in S(\vec{x})}  p(\vec{x} \rightarrow \vec{y}) \phi(\vec{x},t)
\label{master}
\end{align}
where $G(\vec{x})$ is the set of nearest neighbour gas sites of $\vec{x}$ and $S(\vec{x})$ is the set of nearest neighbour solid sites of $\vec{x}$. The transition probabilities are given by
\begin{equation}
\label{trans}
p(\vec{x} \rightarrow \vec{y}) = \frac{1}{d} \times
\begin{cases}
b & \text{ if $\vec{x}$ above $\vec{y}$} \\
1-b & \text{ if $\vec{x}$ below $\vec{y}$} \\
\frac{1}{2} & \text{otherwise}
\end{cases}
\end{equation}
with $d$ the dimension of the lattice and $b \in [0,1]$ a bias. Equations (\ref{master}) and (\ref{trans}) define a discrete equivalent of the advection diffusion equation \cite{Kamp}.

At each time step, the strength of each surface site is reduced in magnitude by the particle density it absorbs. This corrosion rule, together with equation (\ref{master}), and the initial distribution of solid site strengths, defines our model.  Throughout this paper we will assume that the initial strength of each site is a random variable uniform on $[(1-\xi)S, (1+\xi)S]$, and independent of the strength values of its neighbours. The number $S$ is mean site strength, and $\xi$ is referred to as the ``noise''. The variance of site strength is $(\xi S)^2 /3$. The corrosive damage per unit particle density may be varied by altering $S$.

\section{Numerical Results}
\label{numResults}

Equation (\ref{master}) must be solved numerically because of the complexity of the surface once a few surface sites have been destroyed. We consider a finite system and impose lateral periodic boundary conditions, and set the particle density at a fixed distance, $h_G$, above the highest point of surface equal to a constant $\rho >0$. The height $h_G$ is chosen sufficiently large that further increase would not affect the particle density near the surface.

\subsection{One dimensional surfaces}

We consider an initially level surface so that prior to the destruction of the first surface site, the particle density is equal along the interface. In the absence of any noise ($\xi = 0$), all surface sites are simultaneously destroyed, preserving the flat interface, one unit lower. Due to the definition of the height, $h_G$, of the gas domain, the equilibrium particle density at each point of the interface depends only on its level relative to other interface sites. Therefore the corrosion process wears away a flat surface at a constant rate over time.

\begin{figure}
\includegraphics[width=8 cm]{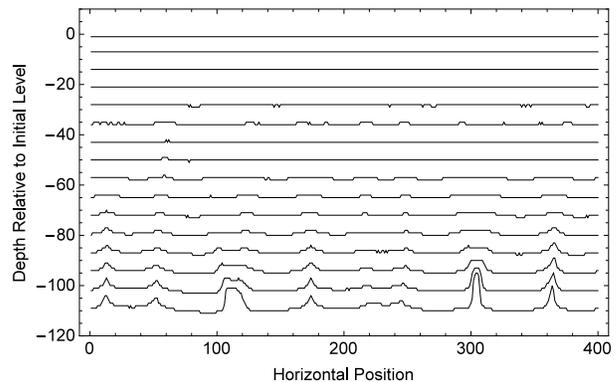}
\caption{ The early evolution of surface fluctuations in the $d=2$ lattice model. The rock surface depth, relative to its starting value, is plotted at 1000 time step intervals. The model parameters in this case are $S=10, \xi = 0.05, b=0.58, p_c=0.025, \rho=1$.    \label{earlyPitEvo}}
\end{figure}

\begin{figure}
\includegraphics[width=8 cm]{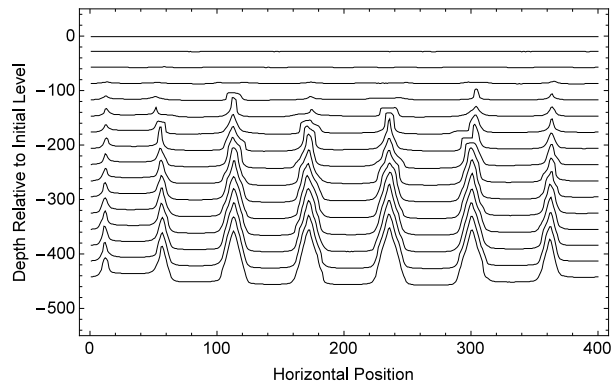}
\caption{ The spontaneous evolution of a regular arrangement of pits in the $d=2$ lattice model. The interface sequence is drawn from the same simulation as Figure \ref{earlyPitEvo} but  plotted at 4000 time step intervals.   \label{spontPitEvo}}
\end{figure}

In the case $\xi>0$, small pits in the surface appear, as observed in metals \cite{Bur04}. Their evolution depends on two effects. First, sites at their edges will have two faces exposed to the gas, and therefore erode more quickly, widening the pit and smoothing the surface. The second effect arises from the on-surface bias of the gas particles which, provided $p_c$ is sufficiently small, causes particles to visit the bases of the pits more frequently, deepening them. Together, these effects cause small fluctuations in the interface to evolve toward larger, smoother pits. The early stages of this process are illustrated in Figure \ref{earlyPitEvo}. These early fluctuations in surface depth lack order. Over a longer period, for example in Figure \ref{spontPitEvo}, the surface evolves toward a stable state characterized by regular trapezoidal pits.

If $p_c$ is too large then the pit deepening effect will vanish because particles are not able to sufficiently explore the surface and find its deepest parts before they are absorbed. This gives rise to a depletion zone in the particle density above the surface.

\begin{figure}
\includegraphics[width=8 cm]{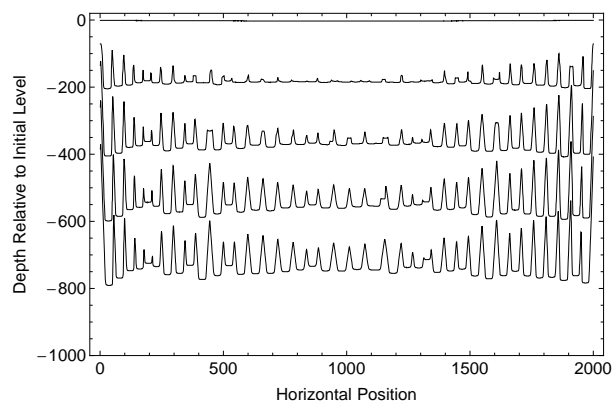}
\caption{ The spontaneous evolution of a regular arrangement of pits in the $d=2$ lattice model with variable corrosion probability. The rock surface depth, relative to its starting value, is plotted at 20,000 time step intervals. The model parameters in this case are $S=10, \xi = 0.05, b=0.6, \rho=1$ with $p_c$ varying piecewise linearly from $p_c=0.01$ at the boundaries to $p_c=0.05$ in the centre of the system.    \label{pcVariation}}
\end{figure}

In Figure \ref{pcVariation} we show the results of varying $p_c$ continuously across the system so that a distance $x$ from the left boundary of a system of width $W$, the corrosion probability follows a triangular distribution: $p_c(x) = \max(p_c) + (\max(p_c)-\min(p_c))\mid 2x-W \mid/W$ with its maximum at $x=W/2$. With lower corrosion probability particles are less likely to be absorbed by pit walls and more likely to drift down to corrode the base, resulting in deeper and narrower structures.

The magnitude of the noise parameter, $\xi$, influences the early stages of the process and although it does not influence the gradient of pit walls, larger noise values roughen the regular trapezoidal formations, and can create narrower pits which later merge. A smaller noise value increases the time taken for fluctuations to develop from the flat surface, and for given values of $b, p_c$ there appears to be a critical value of $\xi$ below which pits fail to emerge. In this work we will focus on the case where $\xi$ is sufficiently large for pits to develop but not large enough to create fluctuations which disrupt regular formations.

\subsection{Dynamics of Pit Depth}

\begin{figure}
\includegraphics[width=7 cm]{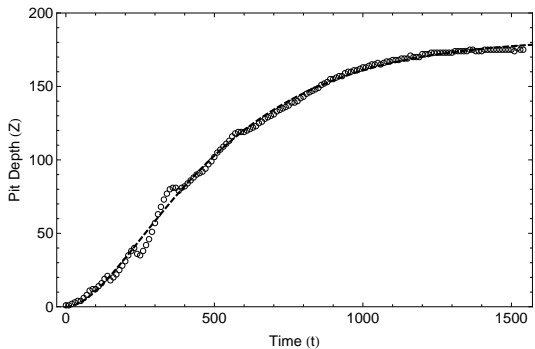}
\caption{ Pit depth (measured by difference between maximum and minimum height of surface) in the $d=2$ lattice model. The variable $t$ is a linear function of the number, $n$, of simulation steps: $t=10^{-2} n-60$, chosen to produce results consistent with field observations of tafoni growth rates \cite{Sun11} measured in years.  The model parameters in this case are $S=10, \xi = 0.05, b=0.58, p_c=0.025, \rho=1$ as in Figure \ref{spontPitEvo}.  Also shown as a dashed line is a fit to the growth curve (\ref{Growth}) with parameter values $n=1$, $\beta=0.0028$ and $Z_c=183$. \label{sCurve}}
\end{figure}

From Figures \ref{earlyPitEvo} and \ref{spontPitEvo} we see that at least $10^4$ time steps pass before the initially flat surface begins to develop sufficient fluctuations for pit development to begin. These ``proto-pits'' initially deepen slowly, and then accelerate before reaching their equilibrium depth, as shown in Figure \ref{sCurve}.
This behaviour matches field observations of tafoni growth rates \cite{Sun11,Turk04,Nor02}. To provide a quantitative comparison to the geomorphology literature, in Figure \ref{sCurve} we also have a graph of the following phenomenological growth curve proposed by Sunamura and Aoki (2011) \cite{Sun11}
\begin{equation}
Z(t) = Z_c \left[ 1- (n+1) e^{-\beta t} + n e^{-(1+n^{-1})\beta t} \right]
\label{Growth}
\end{equation}
where $Z(t)$ is pit depth at time $t$ (years), $Z_c$ is final depth, and $n, \beta$ are physically motivated fitting constants. We have linearly scaled and shifted our simulation time for consistency with their data, selected a value $n=1$ (matching the order of magnitude of their choice) and used $\beta, Z_c$ as fitting parameters. We see that our simulation results provide a close fit to the phenomenological curve, and note that $\beta$ falls within the range of their estimates. Moreover our model gives some insight into the ``feedback'' mechanisms behind the growth process \cite{Sun11,Turk04,Nor02}. We have shown that growth is slow at first while pits take time to emerge from random fluctuations, it then accelerates as pits funnel particles into their bases, evolving toward a trapezoidal equilibrium which finally ceases to deepen due to the presence of neighboring pits.

\subsection{Two dimensional surfaces}

\begin{figure}
\includegraphics[width=7 cm]{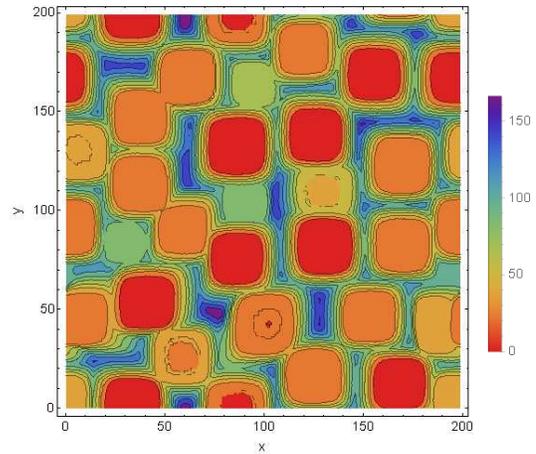}
\caption{ Contour plot of spontaneously evolved pits in the $d=3$ lattice model after 25,000 time steps. The model parameters in this case are $S=5, \xi = 0.2, b=0.6, p_c=0.02$.  \label{contour}}
\end{figure}

\begin{figure}
\includegraphics[width=7 cm]{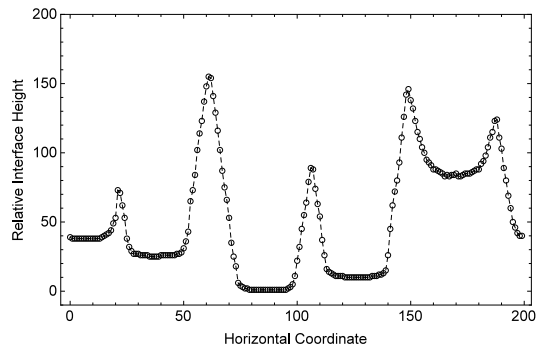}
\caption{ Cross section through surface in the $d=3$ lattice model. The model parameters in this case are $S=5, \xi = 0.2, b=0.6, p_c=0.02$ (identical to Figure \ref{contour}).
\label{cross}}
\end{figure}

\begin{figure}
\includegraphics[width=6.5 cm]{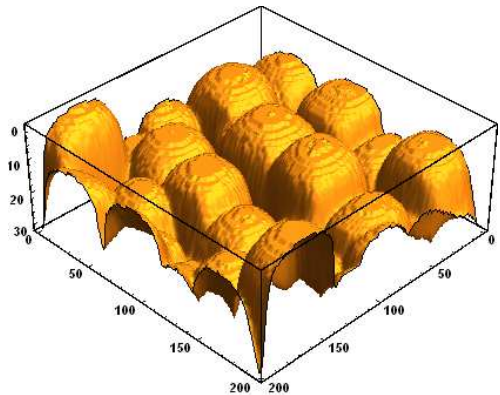}
\caption{ Underside view of spontaneously evolved pits in the $d=3$ lattice model. The model parameters in this case are $S=5, \xi = 0.2, b=0.55, p_c=0.02$. Note that pits are wider and shallower compared to Figures \ref{contour} and \ref{cross} due to lower on-surface bias.
\label{under}}
\end{figure}

When the corrosive particles are incident on a two dimensional interface we again find that early random fluctuations in the interface due to differential site strength evolve toward regular pit structures. As Figure \ref{contour} shows the regularity in spatial arrangement is less pronounced than for the one dimensional interface. However, a cross section through the surface (Figure \ref{cross}) shows  very similar trapezoidal pit shapes are present. As with the one dimensional case, a lower on-surface bias (see Figure \ref{under}) produces wider, shallower pits. This effect appears to be independent of dimension.

\section{Corrosion Gradient Analysis}

We now analyse the corrosion process when the corrosion rate, equivalent to particle density adjacent to the pit walls, is defined externally to the model. The complex interactions between particle flow and surface morphology are replaced with a simple functional relationship between corrosion rate and depth. The relationship is determined by observing the particle density in a spontaneously formed pit in the full model. Figure \ref{wallDen} shows that adjacent to the upper walls the density is approximately constant and increases exponentially toward the base. Also shown is a fitted function of the form $A + B \times \beta^h$. Since this functional form accurately captures the particle density adjacent to the wall and because the evolution of pit shape depends only on this and the corrosion probability $p_c$, then we can expect to at least qualitatively capture the evolution of single pits.

\subsection{Definition of Corrosion Gradient Model}

\begin{figure}
\includegraphics[width=8 cm]{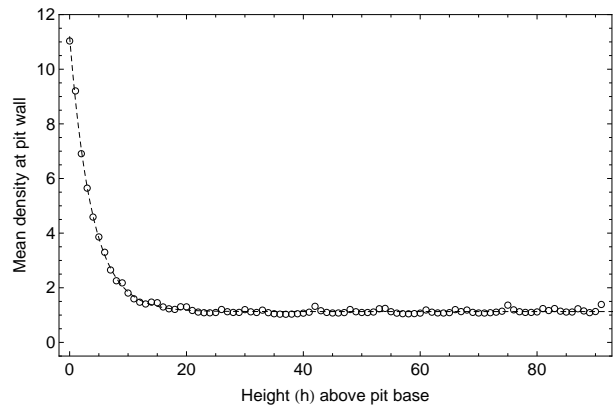}
\caption{ Circles show mean particle density in sites adjacent to pit wall versus height above pit base for the central pit in Figure \ref{spontPitEvo} after $2 \times 10^5$ time steps. The model parameters are $S=10, \xi = 0.05, b=0.58, p_c=0.025, \rho=1$. The dashed line shows the function $A + B \times \beta^h$  where $A = 1.1, B=10.1, \beta = 0.77$.
\label{wallDen}}
\end{figure}

We define a \emph{basin} or \emph{pit} to be single local minimum in a surface (or a line or plane of minima) together with the set of surrounding points which may be connected to the minimum by surface trajectories which do not pass through any maxima or saddle points. We will assume that the corrosion rate at a given point depends only on the height of that point relative to the minimum of the basin to which it belongs. We refer to this as a ``corrosion gradient model''. We define corrosion rate $r(h)$ per site face at relative height $h$ to be:
\begin{equation}
r(h) = 1 + \alpha \beta^h
\label{crate}
\end{equation}
with $\beta \in [0,1]$ and $\alpha>0$. This simplification of the fitting form used in Figure \ref{wallDen} is physically justified since corrosion rate may be rescaled by an overall constant by adjusting initial site strength. We let pits evolve in a similar way to the full model: at each time step the strength of exposed site $\vec{x}$ is reduced by:
\begin{equation}
\Delta S (\vec{x}) = \sum_{\vec{y} \in G(\vec{x})} r[h(\vec{x})]
\end{equation}
where $h(\vec{x})$ is the height of site $\vec{x}$ relative to the minimum of its basin. As in the full model, all solid sites possess an initial strength uniform on $[(1-\xi)S,(1+\xi)S]$ for some choice of $S$ and $\xi < 1$, and are removed from the solid immediately their strength becomes negative.

\begin{figure}
\includegraphics[width=8 cm]{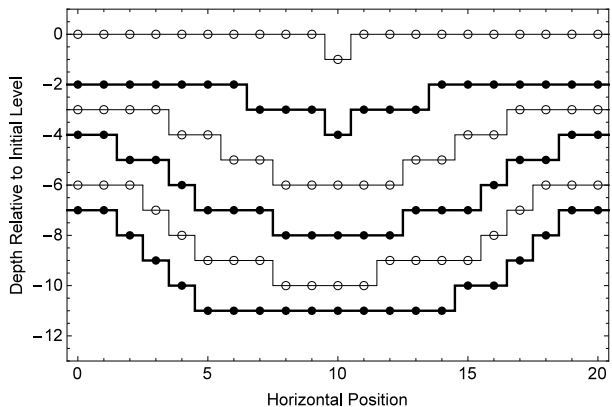}
\caption{ Early stage evolution of a pit in the corrosion gradient model with $\alpha=1, \beta=0.5, S=50, \xi=0.01$. Here the surface is recorded after every 20 successive changes to the boundary.
\label{corroGrad}}
\end{figure}

In Figure \ref{corroGrad} we show a series of detailed snapshots of a surface which begins with a single site removed. As the pit becomes deeper, its base becomes wider and flatter, and a series of \emph{shelves} are formed. Within a given shelf, sites that are closer to the centre of the pit will be weaker because they have been exposed to corrosion for a longer period. The sites comprising a shelf therefore disappear in sequence radially outwards from the centre of the pit.

\subsection{Continuum evolution equation}

In the absence of randomness in the initial strengths of sites, the evolution of the surface is an entirely deterministic process. In this case numerical experiments show that the system finds a stable cycle of surface states. If the initial strengths of sites have nonzero variance ($\xi>0$) then the evolution of the surface is a stochastic process.  In this case the system finds similar, but transient, orbits whose persistence time is greater for smaller noise. By considering the life cycle of a typical site, we will now derive an approximate evolution equation which is able to capture the form of these steady states.

\begin{figure}
\includegraphics[width=6 cm]{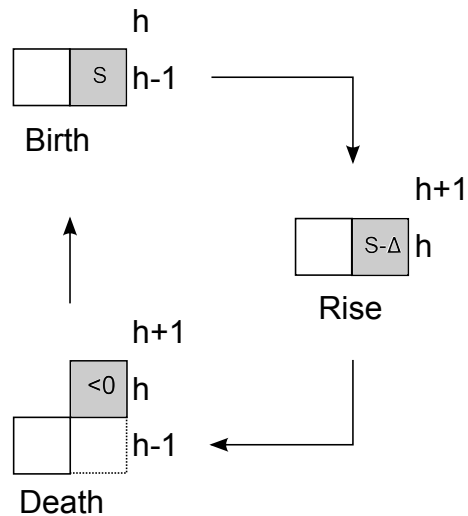}
\caption{ The life cycle of a site (shaded) near the base of the pit. The site is first exposed when $\Delta H =0$, later its relative height increases. Its left neighbor will be destroyed before it.
\label{cycle}}
\end{figure}

We define the random variable  $H(x,t)$ to be the height, relative to pit base, of the uppermost face of the highest surface site at position $x$ at time $t$. Note that the lowest site or sites in the surface have $H=0$ and are exposed to corrosion rate $r(0)=1+\alpha$ on their upper face and cannot have any other faces exposed. If the highest site at position $x$ has a side face and an upper face exposed then it will be exposed to corrosion rate $2+\alpha(\beta^{H-1} + \beta^H)$. We now define
\begin{equation}
\eta(x,t) := \mathbf{E}[H(x,t)].
\end{equation}
We will consider the case where the system evolves from an initial state where there is a unique lowest site in the surface, and we will define its position to be the origin of coordinates; $x=0$. Without loss of generality we consider the shape of the wall which lies to the right ($x>0$) of the base so that $H(x,t) \ge H(x-1,t)$. For given $x$, we define  $ \Delta H (x,t) := H(x,t)-H(x-1,t)$ and note that the number of exposed faces at time $t$ is given by $1+ \Delta H(x,t)$. We also define the discrete derivative of $\eta(x,t)$ with respect to $x$:
\begin{equation}
\eta'(x,t) := \eta(x,t)-\eta(x-1,t).
\end{equation}
We first consider the evolution of $\eta(x,t)$ when $\eta'(x,t)$ is small and $\Delta H \in \{0,1\}$. Simulation results show that this is almost always the case provided we are sufficiently near the base of the pit. Under these conditions $\eta'(x,t)$ is equal to the expected time that the surface site at position $x$ has two faces exposed, and $1-\eta'(x,t)$ is the expected time that it has only one. Because the shelves of sites which form the low levels of the pit are destroyed in sequence radially out from the centre, each surface site must begin life with only its upper surface exposed and end life with two exposed faces. At some point during this lifespan the lowest site in the pit will be destroyed increasing the relative height of all other sites. This cycle of events is illustrated in figure \ref{cycle}.
Taking a weighted sum of the corrosion rates in the two possible $\Delta H$ states we arrive at the following approximate expression for the expected magnitude, $\Delta S$, of the change in surface strength per time step at $x$:
\begin{align}
\nonumber
\mathbf{E}[\Delta S] &\approx [1+\alpha \beta^\eta](1-\eta') + [2 + \alpha( \beta^\eta + \beta^{\eta+1}) ] \eta' \\
&= 1+ \alpha \beta^\eta + (1+\alpha \beta^{\eta+1}) \eta'.
\label{DS}
\end{align}
Away from the base of the pit where $\eta \gg 0$ and $\Delta H$ can take larger values, it is no longer the case that $\eta'(x,t)$ is the expected time for which two faces are exposed, so our derivation ceases to be valid. In this case we may write down a less sophisticated approximation for $\mathbf{E}[\Delta S]$ which does not require $\Delta H \in \{0,1\}$, but ignores the subtleties associated with the cycle illustrated in Figure \ref{cycle}. Since the expected number of exposed faces at position $x$ is $1+\eta'(x,t)$, then by neglecting differences in corrosion rates between the various exposed faces in position $x$, we have $\mathbf{E}[\Delta S] \approx (1+\eta')(1+\alpha \beta^\eta)$. This differs from our original approximation (\ref{DS}) by a quantity exponentially decaying with $\eta$. On the grounds that the $h$ dependent term in $r(h)$ is more significant near the pit base we take equation (\ref{DS}) as our universal approximation for $\mathbf{E}[\Delta S]$.

In order to derive an expression for the discrete time derivative $\dot{\eta}(x,t) := \eta(x,t)-\eta(x,t-1)$ we must take account of a subtle but important correction which arises from the discrete nature of the model. Because the final change in site strength before a surface site is destroyed will, if $\xi>0$, certainly make the strength negative, then solid sites can absorb more units of corrosion than their initial strength value. They therefore have an \emph{effective} strength in excess of their initial strength. If the final change in site strength has magnitude $\Delta S$, then the remaining strength will be approximately uniform on $[0,\Delta S]$ and therefore the expected effective strength of the site will be $S+\tfrac{1}{2} \Delta S$. For sites at level $H=0$,  $\tfrac{1}{2}\Delta S = \tfrac{1}{2}(1 + \alpha) := \epsilon_1$ and for all others $\tfrac{1}{2}\Delta S \approx  \tfrac{1}{2}[2 + \alpha(\beta^\eta + \beta^{\eta+1})] := \epsilon_2$. For a site with $\eta(x)>0$ and $x>0$ then $\eta$ is increased by the corrosion of the pit base, and decreased by corrosion events at position $x$ so that provided $S \gg 1+\alpha$ then for $x>0$
\begin{align}
\dot{\eta} & \approx \frac{1+\alpha}{S+\epsilon_1} - \frac{\mathbf{E}[\Delta S]}{S+\epsilon_2} \\
\label{inter}
& \approx \frac{1+\alpha}{S} - \frac{\mathbf{E}[\Delta S]}{S} + \frac{1+\alpha}{S^2} (\epsilon_2-\epsilon_1) \\
\label{evoPde}
&= \frac{1}{S}\left[ \kappa_0 - \kappa_1 \beta^\eta - (1+\alpha \beta^{\eta+1}) \eta'\right],
\end{align}
where we have defined two constants
\begin{align}
\kappa_0 &= \alpha + \frac{(1+\alpha)(1-\alpha)}{2S} \\
\kappa_1 &= \alpha \left(1 - \frac{(1+\beta)(1+\alpha)}{2S}\right),
\end{align}
and made use of the approximation $\mathbf{E}[\Delta S] \approx 1+\alpha$ in the order $S^{-2}$ term in order to obtain the intermediate equation (\ref{inter}). This relationship holds in equilibrium (when $\dot{\eta}=0$) because all positions must corrode the same rate. Out of equilibrium we are ignoring a correction of order $\mathcal{O}(S^{-2})$ to the time derivative.   The condition $x>0$ for the validity of equation (\ref{evoPde}) is an important one and arises because the site at $x=0$ forming the base of the pit must change in strength by $\Delta S = 1+\alpha$ at every time step and has effective strength $S+\epsilon_1$ so that $\dot{\eta}(0,t)=0$.

\begin{figure}
\includegraphics[width=8 cm]{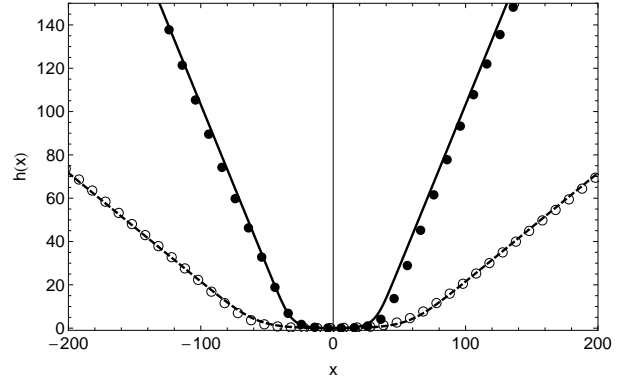}
\caption{ Pit profiles in the corrosion gradient model, together with steady state solutions to the continuum evolution equation. Open circles show the case $\alpha = 0.5, \beta = 0.8, S=200$ and closed circles $\alpha = 1.5, \beta =0.8, S=200$. In both cases the simulation results were obtained with $\xi=0.01$.
\label{steadyStates}}
\end{figure}

Equation (\ref{evoPde}) is a difference equation in two variables and can only be solved numerically. However we can extract analytical information if we interpret $\eta(x,t)$ as a function of continuous time and space variables $x,t$ so that (\ref{evoPde}) becomes a first order partial differential equation.  Since $x=0$ is the deepest point of the pit then $\eta(0,t)=0$ for all $t \ge 0$ and we may find the steady state analytically, subject to this boundary condition, in implicit form:
\begin{equation}
\eta(x) + \left(\frac{\kappa_1+\alpha \beta \kappa_0 }{\kappa_1 \ln \beta}\right) \ln \left[\frac{\kappa_0-\kappa_1}{\kappa_0-\kappa_1 \beta^{\eta(x)}}\right] = \kappa_0 x.
\label{hss}
\end{equation}
Two examples of such steady states are plotted in Figure \ref{steadyStates} along with profiles obtained by simulating the corrosion gradient model. Figure \ref{steadyStates} shows how the steady states of our approximate evolution equation closely match the simulation results, and that the corrosion gradient model generates pits with the same trapezoidal shape found in the full model.
From equation (\ref{hss}) we see that the gradient of the pit wall tends, for large $x$ to $\kappa_0$, which for large $S$ is approximately equal to $\alpha$. In section \ref{trapFun} we will show how this gradient may be related to the parameters of the full model, by viewing each trapezoidal pit as a funnel which concentrates particles as they descend.

\subsection{Pit Widths}

\begin{figure}
\includegraphics[width=8 cm]{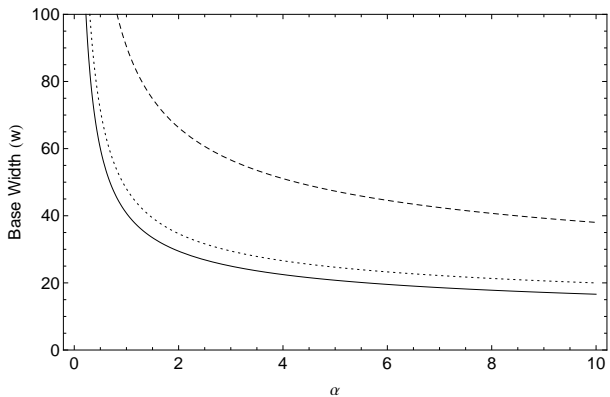}
\caption{ Width of pit base versus $\alpha$ (equivalent to wall gradient in the limit $S \rightarrow \infty$). Parameter values are $\beta=0.8, S=500$ (dotted) $\beta=0.8, S=200$ (solid) and $\beta=0.9, S=200$ (dashed).
\label{widths}}
\end{figure}

By making use of our implicit solution for the steady state pit profile we may derive an analytical expression for the width of its base. We define the edge of the pit as the solution to $\eta^{(3)}(x)=0$, which is the inflection point in the gradient of the wall. An implicit expression for $\eta^{(3)}(x)$ in terms of $\eta(x)$ may be obtained by repeatedly differentiating the steady state differential equation for $\eta(x)$. The condition $\eta^{(3)}(x)=0$ then reduces to a third order polynomial in $\beta^\eta$, having solution:
\begin{equation}
\beta^\eta = \frac{1 + \alpha \beta -\sqrt{1 + \alpha \beta + \alpha^2 \beta^2}}{\alpha \beta} + \mathcal{O}\left(\frac{1}{S}\right)
\end{equation}
Substitution into equation (\ref{hss}) gives an analytic expression for the width, $w$, of the pit base, which we provide here to lowest order in $S$:
\begin{align}
\nonumber
w &:= \frac{1}{\alpha \ln \beta} \ln \left[ \frac{1 + \alpha \beta -\sqrt{1 + \alpha \beta + \alpha^2 \beta^2}}{\alpha \beta} \right] \\
\nonumber
&+ \frac{1+ \alpha \beta}{\alpha \ln \beta} \ln \left[\frac{(1 + \alpha)\left[1 + \sqrt{1 + \alpha \beta + \alpha^2 \beta^2}\right] }{2 \alpha  S} \right] \\
&+ \mathcal{O} \left(\frac{\ln S}{S} \right)
\label{w}
\end{align}
In the limit $\alpha \rightarrow 0$ both terms diverge in magnitude but with opposite signs, the second term being positive with a higher order divergence. Therefore the pit width tends to infinity as $\alpha$ (the wall gradient) tends to zero. Narrower based pits will therefore have steeper walls. As $S \rightarrow \infty$ the second term possesses a divergence $\propto \ln S$ implying that harder surfaces produce wider pits. Both terms diverge as $\beta \rightarrow 1$. This limit is equivalent to the limit of zero downward bias. Figure \ref{widths} illustrates these effects.

\subsection{Formation Dynamics}

\begin{figure}
\includegraphics[width=8 cm]{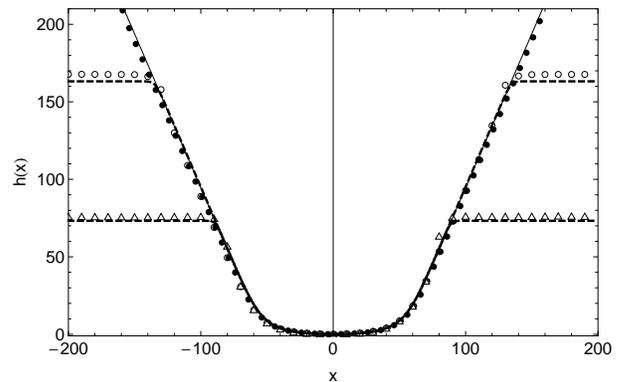}
\caption{Simulated pit profiles at times $t = 5.53 \times 10^3$ (triangles) $t=1.01 \times 10^4$ (open circles) and $t=1.39 \times 10^4 \}$ closed circles.  Parameter values are $\alpha = 2.0, \beta=0.9, S=100, \xi=0.01$. The thick dashed lines give numerical solutions to equation (\ref{evoPde}) at the corresponding times, whereas the continuous line gives the steady state.
\label{dynam}}
\end{figure}

We now turn to the dynamical process by which pits are formed, which is described approximately by equation (\ref{evoPde}). We will interpret this as a partial differential equation in $x$ and $t$, but note that because of the condition $\dot{\eta}(0,t)=0$ it is not analytically tractable. We solve the equation numerically for $x \ge 0$ using the method of lines \cite{Sch09} from an initial condition $\eta(x,0)=0$. In Figure \ref{dynam} we see that as the pit deepens the internal structure which has already formed is preserved.
Whilst in this example the pit can continue to deepen indefinitely, in the full model the presence of spontaneously formed neighboring pits limits their depth.

\section{Trapezoidal Funnel Effect}
\label{trapFun}

We may find an approximate relationship between the parameter $\alpha$ of the corrosion gradient analysis and the parameters $b, p_c$ by considering the stability of a trapezoidal pit in the surface of the full model. Because particles are biased downwards, the pit acts like a funnel which concentrates the particles into a narrower space as they descend. However, the funnel has absorbing sides which counteract this effect. If these two effects are not in balance at the mouth of the funnel, then they will tend toward a state of balance at lower levels because increasing particle concentration leads to an increase in absorbtion rate.  However, changes in corrosion rate with depth will distort the constant wall gradient over time making the trapezoidal geometry unstable. Therefore, in order for a trapezoidal pit to be stable, funnelling and absorption must be in balance at the mouth of the pit.

Given that we expect particle density near the wall of a stable trapezoidal pit to be equal at all levels, then if  particle density is approximately constant across the pit mouth, it must remain so a lower levels. This observation leads to an analytical approximation for stable pit gradient, derived using simple random walks. The expression is approximate because the presence of peaks in the pit structure and discrete steps in the pit wall distort the particle density at the mouth and adjacent to the walls. We will take account of these effects in a more technical but less tractable calculation.

\subsection{Constant density approximation using simple random walk}

The net downward drift of a particle in the vicinity of a sloping pit wall will tend to bring it closer to the wall. If the gradient of the wall is $m$, then in a reference frame with its origin at the wall, but moving so as to remain level with the particle, the particle will appear to have a net velocity toward the wall
\begin{equation}
v = \frac{2b-1}{2m}.
\label{v_wall}
\end{equation}
We assume that $m \ge 1$ so that the horizontal motion in this reference frame is a discrete time simple random walk \cite{Law06}. We will approximate this walk as uncorrelated with
\begin{equation}
\mathbf{P}(\text{no step}) = \frac{m-1}{2m}
\label{still}
\end{equation}
which is the fraction of steps in two dimensions which do not change the horizontal distance between the particle and the wall, accounting for the fact that the wall is comprised of vertical faces. We also assume that the pit is sufficiently wide so that the influence of the opposite wall can be neglected and with effectively infinite depth so that we need not consider the influence of the base.  Letting $a$ be the probability of a move toward the wall, with the probability of remaining still given by (\ref{still}), then the correct net velocity (\ref{v_wall}) is obtained if:
\begin{equation}
a = \frac{2b+m}{4m}.
\label{adrift}
\end{equation}
Let $\phi_k$ be the equilibrium particle density $k$ steps away from the position ($k=0$) immediately adjacent to the wall then for $k>0$:
\begin{equation}
(2b+m) \phi_{k+1} - 2(m+1)\phi_{k} + (2-2b+m) \phi_{k-1} = 0.
\label{DE}
\end{equation}
A fraction $1/m$ of the sites adjacent to the wall are corner sites since they they are bordered by two wall sites, one to the side and one below. We assume that conditional on a particle being adjacent to the wall, the probability that it occupies a corner site is $1/m$. In this case the probability that a particle which is adjacent to the wall will attempt to jump into it is $(m+2b)/4m$. Given that $p_c$ is the probability of absorption if a particle attempts to step into the wall, then the following boundary condition holds:
\begin{equation}
[m+2(1-b) + (2b+m)p_c]\phi_0 = [2b+m] \phi_1.
\label{BC}
\end{equation}
Since we are treating our pit as having effectively infinite width then, as $k \rightarrow \infty$ the particle density must tend to its value at height $h_G$ (see section \ref{numResults}) above the surface: $\rho=1$. We therefore require that $\lim_{k \rightarrow \infty} \phi_k = 1$. Solving equation (\ref{DE}) under these conditions we find that:
\begin{equation}
\phi_k = 1 + c \left(\frac{1+m(1-2a)}{2a m}\right)^k,
\label{phi}
\end{equation}
where
\begin{equation}
c = \frac{2(2b-1)-(2b+m)p_c}{(2b+m)p_c}.
\end{equation}
We approximate the density profile at the mouth of the pit with its value above the surface. If the pit gradient is stable we expect the equilibrium profile to match this: $\phi_k = 1$ for all $k \ge 0$ so $c=0$. Imposing this condition we find that
\begin{equation}
m(b,p_c) = \frac{2[(2-p_c)b-1]}{p_c}
\label{grad}
\end{equation}
which is the approximate gradient at which the particle density across the pit will remain constant, until the influence of the base or the interaction between the two opposing walls becomes significant.

\subsection{Simulated wall gradient}

\begin{figure}
\includegraphics[width=8 cm]{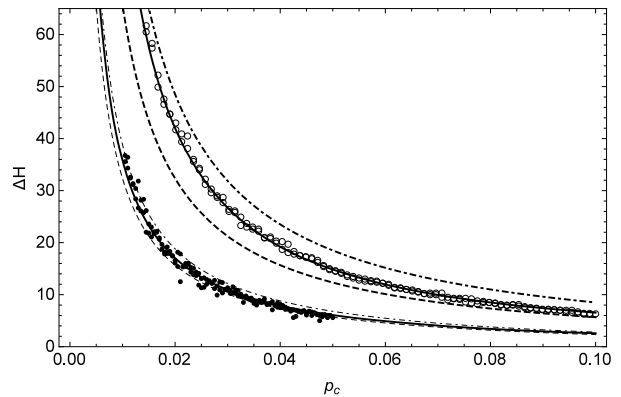}
\caption{ Open circles show estimated wall gradients (defined in main text) versus $p_c$ in a system of length 4000 where $p_c \in [0.01,0.1]$ and $b=0.75$. Dots show estimated wall gradients versus $p_c$ in a system of length 4000 where $p_c \in [0.01,0.05]$ and $b=0.6$. The dot-dashed lines are graphs of $m(b,p_c)$ defined in equation (\ref{grad}) for $b=0.75$ (thick line) and $b=0.6$ (thin line). The dashed lines are graphs of $\gamma(b,p_c)$ (equation (\ref{gamma})) for $b=0.75$ (thick line) and $b=0.6$ (thin line). The solid lines are the sloping wall approximations described in section \ref{exactSlope}.
\label{wallGrad}}
\end{figure}

We have estimated the wall gradient as a function of $p_c$ for various fixed bias values by simulating a surface in a $d=2$ system where the the corrosion probability varies slowly and continuously with position. In order to relate wall gradient to position we compute the set, $\mathcal{D}_1(x,N)$, of finite difference first derivatives in a window of width $2N+1$ for some $N >0$ centred on position $x$
\begin{equation}
\mathcal{D}_1(x,N) := \left\{ H(k)-H(k-1) \right\}_{k=x-N}^{x+N}.
\end{equation}
The constant gradient sections of wall are the steepest parts of the surface, and provided the window is significantly wider than the pit base then each window will contain such a section. The wall gradient may then be estimated as the mean of the largest $n$ elements of $\mathcal{D}_1(x,N)$ with $n$ small enough so that all $n$ elements belong to a constant gradient section. In Figure \ref{wallGrad} this process has been used to estimate wall gradients with $N=25$ and $n=10$ for $b \in \{0.6,0.75\}$. Also shown in Figure \ref{wallGrad} are graphs of $m(b,p_c)$ defined in equation (\ref{grad}), which is our estimated gradient if the particle density were equal to unity across the mouth of the pit. We see that whilst (\ref{grad}) is a close approximation when $b=0.6$, the approximation is poorer when $b=0.75$. We now address this point.

\subsection{Approximation using exact equilibrium density profile near a sloping wall}
\label{exactSlope}

\begin{figure}
\includegraphics[width=8 cm]{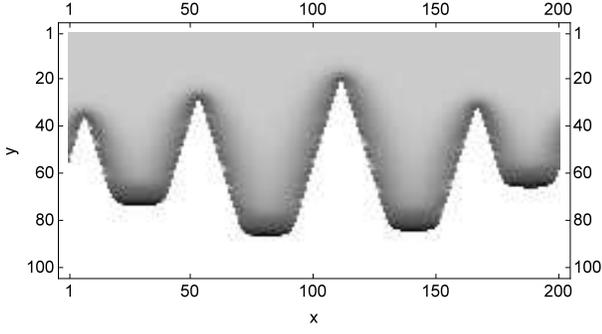}
\caption{ Plot of the particle density field where higher density is darker gray. Parameter values are $p_c=0.1, b=0.65,S=10,\xi=0.05$.
\label{densityPlot}}
\end{figure}

The presence of peaks in the pitted surface distorts the particle field at the mouths of the pits. Regions of increased density appear immediately above the peaks where particles collect before entering the pit (Figure \ref{densityPlot}).
Also, the discrete nature of the surface creates steps in the pit wall above which the particle density is increased. We ignored these effects in our simple analysis by assuming that particle density is constant across the pit mouth, and that all sites adjacent to the pit wall have equal particle density. These approximations can be improved upon using an exact calculation of the equilibrium particle density near the stepped wall.

\begin{figure}
\includegraphics[width=6 cm]{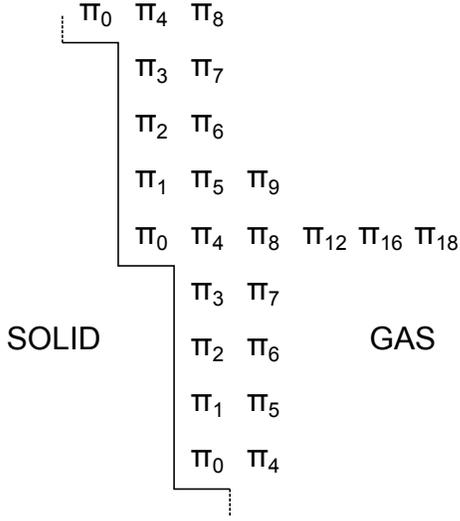}
\caption{ Schematic representation of the equilibrium density distribution adjacent to a discrete sloping wall of gradient $m=4$. The variables $\pi_i$ are particle densities.
\label{discWall}}
\end{figure}

\begin{figure}
\includegraphics[width=8 cm]{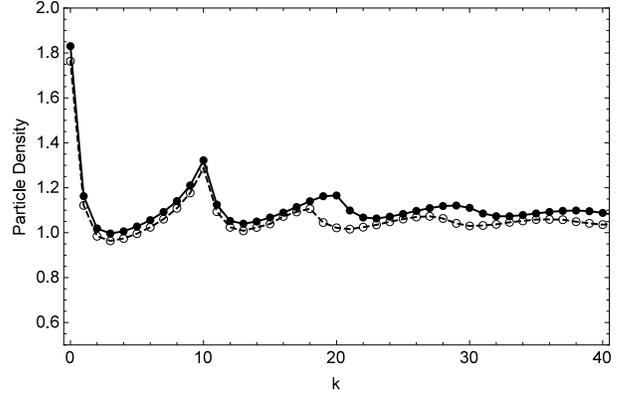}
\caption{ Filled circles with solid line show the exact equilibrium particle density in a column above a step in a discrete sloping wall with gradient $m=10$ and $b=0.75, p_c=0.07$. Open circles with dashed line shows the particle density in a column above a step in a spontaneously formed sloping wall in the full model with  $b=0.75, p_c=0.07$.
\label{piSol}}
\end{figure}

We consider a single step in an infinite sloping wall with gradient $m \in \{1,2,3, \ldots \}$. Let the particle density in the site immediately above the step be $\pi_0$, and the density $k$ steps above that be $\pi_k$. Figure \ref{discWall} illustrates the case where $m=4$.
If the particle density is in equilibrium, then the column of sites above every step must have an identical density profile, as shown in Figure \ref{discWall}. In sites not adjacent to the wall, having indices $k \ge m$, we have
\begin{equation}
\pi_{k+m} + 2b \pi_{k+1} - 4 \pi_k + 2(1-b) \pi_{k-1} + \pi_{k-m} =0.
\end{equation}
The general solution to this equation \cite{Gol86} may be written in terms of the roots, $\{\lambda_1, \ldots, \lambda_{2m}\}$ of the characteristic polynomial
\begin{equation}
\lambda^{2m} + 2b \lambda^{m+1} - 4 \lambda^m + 2(1-b) \lambda^{m-1} +1 =0.
\label{poly}
\end{equation}
We will assume the roots to be ordered by absolute value so that $\mid\lambda_{k+1}\mid \ge\mid\lambda_k\mid$.
If $\lim_{k \rightarrow \infty} \pi_k = 1$ then only roots with $\mid \lambda_k \mid \leq 1$, of which there are $m+1$, the largest of which is $\lambda_{m+1}=1$, can contribute to the solution:
\begin{equation}
\pi_n = 1 + \sum_{k=1}^m c_k \lambda_k^n.
\end{equation}
The constants $c_k$ must be found using the boundary conditions for sites adjacent to the wall:
\begin{align}
\label{wallCond}
(3 + p_c) \pi_k - \pi_{k+m} - 2b \pi_{k+1} - 2(1-b) \pi_{k-1} &= 0  \\
[3-2b + (2b+1)p_c] \pi_0 - \pi_m - 2b \pi_1 &= 0
\end{align}
where (\ref{wallCond}) holds for $k \in \{1, \ldots, m-1\}$. An example of this solution is shown in Figure \ref{piSol}.
We note that the solution oscillates with period $m$, matching the period of steps in the wall .

Also shown in Figure \ref{piSol} is the density in a column of sites above a step in a spontaneously formed sloping wall in the full model with the same $b, p_c$ values that were used in the exact solution (the full model was simulated first and the value $m=10$ in the exact solution chosen to match the spontaneously formed wall gradient). Although the average particle density near the wall is much larger than the bulk value ($\rho=1$), the minimum of each oscillation is approximately equal to this value. This suggests that an appropriate condition on the exact solution required to relate $m$ to $b$ and $p_c$ is $\min\{\pi_0, \pi_1, \ldots, \pi_{m-1}\} = 1$. For a given value $b$, we may determine the relationship between $m$ and $p_c$ by finding the value of $p_c$ for which this condition holds. The solid curves in Figure \ref{wallGrad} were obtained by the method, and we see that the gradient estimates are accurate. We note however that it is not possible to write down an analytical relation expression for $m$.


\subsection{Connection to corrosion gradient analysis}

Our theoretical estimates of the wall gradient provide a link between the full model and the corrosion gradient model. In this latter model we found that the wall gradient is approximately equal (with corrections of order $1/S$) to the constant $\alpha$ in the corrosion rate function $r(h) = 1 + \alpha \beta^h$  (equation (\ref{crate})). The trapezoidal funnel effect also explains why the particle density, and therefore the corrosion rate, is approximately constant adjacent to the pit walls. It remains to interpret the constant $\beta$ in terms of the full model. Although we cannot provide a precise relationship between $\beta, b$ and $p_c$, insight into both wall gradient and $\beta$ may be gained by considering the particle density above the base of the pit. We have argued that particles should neither be concentrated nor depleted as we descend into the pit, so we expect the particle density above the flat base to be independent of the pit depth and therefore to take a similar form to the equilibrium particle density above a flat, partially absorbing surface. Considering only the vertical motion of a particle above such a surface in the full model, and letting $\psi_k$ be the equilibrium density $k$ steps above the site adjacent to the surface we have:
\begin{align}
\psi_k &= b \psi_{k+1} + (1-b) \psi_{k-1} \\
b \psi_1 &= [1-(1-p_c)b] \psi_0,
\end{align}
with $\lim_{k \rightarrow \infty} \psi_k = 1$. Solving for $\psi_k$ we find that:
\begin{equation}
\psi_k = 1 + \left[\frac{(2-p_c)b-1}{b p_c} \right] \left(\frac{1-b}{b}\right)^k.
\label{psi}
\end{equation}
We note the similarity between the wall gradient approximation (\ref{grad}) and the coefficient of the exponential term in (\ref{psi}) which we define as a new function
\begin{equation}
\gamma(b,p_c) = \frac{(2-p_c)b-1}{b p_c}.
\label{gamma}
\end{equation}
In the limit $b \rightarrow \tfrac{1}{2}$ the functions $\gamma(b,p_c)$ and $m(b,p_c)$ become identical. The function $\gamma(b,p_c)$ is graphed in Figure \ref{wallGrad} and we see that although it underestimates the wall gradient, the error is of similar magnitude to equation (\ref{grad}). This suggests equation (\ref{psi}) as a crude approximation to the corrosion rate function (\ref{crate}) so that $\beta \approx (1-b)/b$. For the parameter values used in Figure \ref{wallDen} we found by regression that $\beta \approx 0.77$. In that case the bias was, $b=0.58$ which gives $(1-b)/b = 0.72$. We have also computed $\beta$ by regression using the density profile in Figure \ref{densityPlot}, finding that $\beta \approx 0.58$. In that case, $b=0.65$ which gives $(1-b)/b = 0.54$.

\section{Discussion and Conclusion}

We have introduced a simple lattice model of surface damage by incident particles in which regular pits spontaneously form in one and two dimensional surfaces as a consequence of convection of corrosive material toward the surface, along with some fixed probability of absorbtion at each contact. Such regular pit formations have been observed in rocks \cite{Must82,Hui04,Sun11,McC79} and also in metals \cite{Pid08,Bur04}. Our model provides a highly simplified view of reality which nevertheless provides some insight into field observations \cite{Sun11} of the growth rate of pits. We also note that the convection of eroding particles into the deeper parts of pits shares an important feature with other explanations of pit formation in rocks \cite{Hui04,Mott94}: corrosion rates are greater deeper in the rock core. Our corrosion gradient approach provides a natural framework for incorporating more sophisticated modelling approaches through the corrosion rate function, which could be calibrated to capture corrosion processes other than convection and absorption. We have also been able to analytically relate the geometry of pits to the corrosion rate function, and to the hardness of the surface.

An analytical theory for the spontaneous regular arrangement of pits within our model remains to be found. The fluctuations, seen in Figure \ref{earlyPitEvo}, in the early stages of the formation process may be viewed as a superposition of multiple proto-pits most of which are absorbed into other larger pits. Two effects appear important to the arrangement process. First, there is a critical depth at which pits become stable in the sense that a large neighboring pit will not absorb them. Below this depth pits can coagulate. Second, pits appear to be able to migrate small distances across the surface. We conjecture that the combination of migration and coagulation is responsible for regularity in the final arrangement.

\begin{acknowledgments}
James Burridge would like to thank Samia Burridge for carefully reading and discussing this work.
\end{acknowledgments}


\end{document}